\documentclass[%
 aip,
amsmath,amssymb,
 reprint,%
]{revtex4-1}

\usepackage{graphicx}% Include figure files
\usepackage{dcolumn}% Align table columns on decimal point
\usepackage{bm}% bold math
\usepackage{cleveref}
\usepackage[utf8]{inputenc}
\usepackage[T1]{fontenc}
\usepackage{mathptmx}
\usepackage{etoolbox}
\usepackage{xcolor}

\makeatletter
\def\@email#1#2{%
 \endgroup
 \patchcmd{\titleblock@produce}
  {\frontmatter@RRAPformat}
  {\frontmatter@RRAPformat{\produce@RRAP{*#1\href{mailto:#2}{#2}}}\frontmatter@RRAPformat}
  {}{}
}%
\makeatother
\graphicspath{{Figures/}}
\begin{document}

\title{Real-time dynamics with bead-Fourier path integrals. III. Atomistic condensed-phase simulations}

\author{Nathan London}
\author{Mohammad R. Momeni}
\email{mmomenitaheri@umkc.edu}
\date{}
\affiliation{Division of Energy, Matter and Systems, School of Science and Engineering, University of Missouri $-$ Kansas City, Kansas City 64110, MO United States.}

\date{\today}% It is always \today, today,

\begin{abstract}
Accurately accounting for nuclear quantum effects (NQEs), including zero-point energy and nuclear tunneling, is essential for a reliable prediction of the structural, thermodynamic, and dynamical properties of chemical systems, particularly those involving light atoms. However, incorporating NQEs could render simulations of large-scale chemical systems computationally prohibitive, particularly when combined with expensive electronic-structure calculations. One potential avenue for reducing cost is to use bead-Fourier path integrals (BF-PIs). In Paper I\cite{octLondon2025}, we introduced the BF centroid molecular dynamics (BF-CMD) method, in which BF-PIs were used to compute an effective potential for CMD, and in Paper II\cite{BF2}, we introduced BF ring polymer MD (BF-RPMD), taking the BF-PI and using it for approximate real-time dynamics. Both papers featured a series of model systems and were able to reduce the number of beads needed to converge dynamical properties by a factor of 2 to 4, depending on the system. In this work, we report the first implementation and application of the BF-PI framework to atomistic simulations using liquid bulk water and hexagonal ice as test cases. As in our previous papers, we find that adding just a single Fourier component to the path yields a 4- and 8-fold reduction in the number of beads while achieving results that agree with reference PI simulations for liquid water and ice, respectively. We also demonstrate that we can reduce simulation costs by up to 5.5x compared to standard PIs by reducing the number of force calculations and normal-mode transformations.
\end{abstract}

\maketitle

\section{Introduction}

Feynman path integral (PI)-based dynamics methods\cite{julFeynman2010} are commonly used to readily incorporate nuclear quantum effects (NQEs) into standard molecular dynamics (MD) simulations. The NQEs are introduced by creating a number of replicas, or ``beads'', of the particles connected via harmonic springs to form a ring polymer (RP).\cite{julFeynman2010,aprCeperley1995} The spatial distribution of these beads captures quantum nuclear delocalization and fluctuations arising from NQEs, allowing the system to sample configurations that deviate from those predicted by classical MD. While exact equilibrium properties can be calculated using PIMD methods, real-time quantum dynamics can only be approximated using various methods, with RPMD\cite{augCraig2004,febCraig2005,julCraig2005} being the simplest. The replicas of the system only interact with each other through the harmonic springs, not through standard interatomic potentials. This allows for the forces to be calculated separately for each replica, making these PI-based methods scale roughly linearly with the number of beads used to quantize the system (i.e., $\mathcal{O}$ \emph{PN logN}). While this does not pose a barrier to simulations that use analytical force fields (FFs), it does pose significant concerns when more accurate force calculations are used in MD, such as in \emph{ab initio} (AI)MD ($\mathcal{O}$ \emph{PN\textsuperscript{3}}) or high-dimensional neural network potentials (HDNNPs, $\mathcal{O}$ \emph{PN\textsuperscript{2}logN}).

Many methods have been developed to make PI-based dynamics methods more computationally efficient. The first category of these methods aims to reduce the number of beads in the simulations based on different contexts. The ring polymer contraction (RPC) method\cite{julMarkland2008} reduces the number of expensive force calculations. This was originally done in the context of FFs, where the rapidly varying intramolecular forces are evaluated at the full-bead level, while the slowly varying intermolecular forces are calculated on a contracted set of beads and then projected onto the full set.\cite{julMarkland2008} This thus requires identifying the separation between those types of forces. The method has also been extended beyond FFs, i.e., AI-RPC\cite{febMarsalek2016}, where the full bead forces are evaluated using a cheaper electronic structure method while a difference potential is evaluated at a more accurate level on the contracted set of beads. RPC has also been used in the context of NNPs, where two different equivariant transformer NNPs were trained, with one using a short local cutoff radius, and thus larger errors in the energy and forces.\cite{octChen2025} The short-range NNP was then used as the reference potential, with the full NNP used to calculate the difference potential.\cite{octChen2025}.

Another method in this vein is mixed time slicing (MTS).\cite{febSteele2011} MTS uses the simple idea that in a general atomistic simulation, the extent of NQEs for a given type of atom varies, with NQEs typically being more pronounced for atoms with smaller masses. As such, the number of beads needed to fully capture the NQEs of heavier atoms is less than that for lighter ones. In the MTS framework, for atoms with fewer beads, each of their beads will interact with multiple beads of the atoms being described with more beads.\cite{febSteele2011} While this approach can reduce the number of force evaluations for certain interactions, such as bonded interactions between the heavier atoms, non-bonded interactions would typically involve doing a full set of evaluations.\cite{junJohnson2026} This does limit the versatility of the method, where there does not appear to be any benefit to the method in the AI-PIMD context.

The second category of methods involves reevaluating the PI Hamiltonian. The conventional PI framework is derived by taking the Trotter decomposition of the Boltzmann operator to separate the kinetic and potential energy operators of the Hamiltonian to obtain the PIs. Several methods have been developed that include high-order decompositions of the Boltzmann operator. These higher-order schemes introduce the Hessian matrix into the Hamiltonian, which can be a computational bottleneck, but there are ways to approximate the forces through a projection of the Hessian.\cite{novZhang2025} While these higher-order methods have demonstrated computational speedup for calculating thermal properties with PIMD simulations, their use for real-time dynamics is less straightforward. For example, thermostatted (T)-RPMD using the Suzuki-Chin\cite{junSuzuki1995,marChin1997} decomposition showed slower convergence when calculating vibrational density of states.\cite{decKapil2016} While T-RPMD using the Takahashi-Imada\cite{novTakahashi1984} decomposition showed improved convergence, the altered Hamiltonian created blue shifts in the calculated spectra, requiring an ad hoc correction term\cite{sepShepherd2021} that could vary between systems.

Another commonly used method for improving computational efficiency of PI simulations is the PI generalized Langevin equation thermostat (PIGLET) method.\cite{sepCeriotti2010} Using fewer beads to describe the system does not fully account for the full quantum fluctuations of the atoms, and the PIGLET method uses a stochastic thermostat in the form of a GLE thermostat. This thermostat includes auxiliary degrees of freedom (DOFs) the friction parameters for it are carefully tuned to ensure that the reduced number of beads is sampling the same extended phase-space as the full RP would.\cite{sepCeriotti2010} The method works well for calculating equilibrium properties, but suffers in real-time dynamics in a similar manner to the higher-order schemes, where the real-time dynamics are significantly altered and require corrections when calculating properties like vibrational spectra.\cite{marKapil2020} Additionally, the thermostat requires the use of system-specific friction parameters that are not necessarily simple to obtain for a general system.

A far less explored area of PI methods intended to reduce the number of beads needed to represent a system is that of bead-Fourier (BF)-PIs. The method combines the alternative PI framework of Fourier PIs, in which the quantum path is approximated through a Fourier sine series\cite{marDoll1984,julDoll1985}, with the more common bead-based PIs. BF-PI Monte Carlo\cite{febVorontsov-Velyaminov1997} and BF-PIMD\cite{junIvanov2003,julIvanov2005} methods used for model systems show that the addition of a few terms of the Fourier sine series can greatly reduce the number of beads needed to converge thermal equilibrium properties.

Until recently, BF-PIs had been used only in the context of thermal properties and had not been utilized to approximate real-time quantum dynamics. In Paper I\cite{octLondon2025} we introduced the BF-centroid MD (BF-CMD) method, where we used BF-PIs to generate a potential of mean force that was used to propagate real-time dynamics of a classical-like particle. We were able to capture the same potential as using a standard bead PI but with a 4-fold reduction in the number of beads by adding a single Fourier component.\cite{octLondon2025} In Paper II\cite{BF2}, we extended BF-PIMD into approximate real-time dynamics as BF-RPMD. As with BF-CMD, we observed up to a 4-fold reduction in the number of beads with the addition of a single Fourier component.\cite{BF2} Both of these previous papers were limited to one-dimensional model systems. In this work, we report the first implementation and application of the BF-PI framework to atomistic simulations using liquid bulk water and hexagonal ice as test cases. This work thus represents the first use of BF-PIs for condensed-phase atomistic simulations.

The rest of the paper is structured as follows: in Sec.~\ref{sec:Theory} we briefly outline the theory of BF-PIMD and BF-RPMD frameworks. Sec. \ref{sec:Details} discusses the efficient implementation of atomistic BF-PIMD and BF-RPMD methods into our in-house MD code, DL\_POLY Quantum, as well as the simulation details for the systems considered in this work. We discuss the results and performance of atomistic BF-PIMD and BF-RPMD simulations in Sec.~\ref{sec:Results}. Finally, we discuss the future paths for BF-PIs in Sec.~\ref{sec:Conclusion} 

\section{Theory}\label{sec:Theory}

A detailed discussion of the BF-PI framework is presented in Paper II\cite{BF2}. Here we present a brief description of the method for use in atomistic BF-PIMD and BF-RPMD simulations, with a focus on extensions beyond a model system.

In BF-PIs, the imaginary time quantum path is represented using discrete time slices (the
beads in B-PIs) that are then connected with a continuous path described with a Fourier sine series (as in F-PIs)
\cite{febVorontsov-Velyaminov1997}. For a system of $N$ particles quantized using $n$ beads, the BF path associated with the $j^{th}$ bead of the $i^{th}$ atom is defined as,
\begin{equation}
  \mathbf{q}_{i,j}(\xi) = \mathbf{q}_{i,j} + (\mathbf{q}_{i,j+1} - \mathbf{q}_{i,j})\xi + \sum_{k=1}^{k_{\mathrm{max}}}
  %\frac{\sqrt{2}}{k\pi} 
  \mathbf{a}_{i,jk} \sin(k\pi\xi),
  \label{eq:paths}
\end{equation}
where $\mathbf{q}_{i,j}$ is the position of that bead, $\mathbf{a}_{i,jk}$ is the $k^{th}$ Fourier component of the
path, and $\xi$, which takes on a continuous value between 0 and 1, is the distance along the imaginary time path
between beads $j$ and $j+1$.

The BF Hamiltonian for this system is,\cite{febVorontsov-Velyaminov1997,junIvanov2003}
\begin{multline}
  H_{\mathrm{BF}} = \sum_{i=1}^N\left[ \sum_{j=1}^n \left[ \frac{\mathbf{p}_{i,j}^2}{2m_{i,n}} +
    \sum_{k=1}^{k_{\mathrm{max}}} \frac{\mathbf{p}_{i,jk}^2}{2m_{i,k}}\right] \right.
    \\ \left.+ \sum_{j=1}^n\left[ \frac{1}{2}m_{i,n}\omega_{n}^2\left(\left(\mathbf{q}_{i,j+1}-\mathbf{q}_{i,j}\right)^2 +
      \sum_{k=1}^{k_{\mathrm{max}}}\frac{(k\pi)^2}{2}\mathbf{a}_{i,jk}^2\right) \right. \right.
      \\ \left. \left.+ \frac{1}{n} \int_{0}^1 \mathrm{d}\xi\ 
        V\left[\mathbf{q}_{i,j}(\xi)\right]
        \right]
    \right],
  \label{eq:BF-Ham}
\end{multline}
where $\mathbf{p}_{i,j}$ and $\mathbf{p}_{i,jk}$ are the bead and Fourier component momenta, $m_{i}$ is the mass \cite{janParrinello1984a} of the
$i^{th}$ atom and $m_{i,k}$ is the mass of the $k^{th}$ Fourier component, $m_{i,n}=\tfrac{m_{i}}{n}$,
$\omega_{n}=\tfrac{n}{\beta\hbar}$, and $V(\mathbf{q})$ is the external potential. The $ 1/n $ factors in the
Hamiltonian are included due to the use of $ \beta $ instead of $ \beta_{n} $ in the BF-PI partition function.\cite{BF2}
%The use of the implicitly scaled
%Fourier components result in the term describing their harmonic motion to be independent of $k$ \cite{Ivanov2005}.

To perform dynamical simulations, the equations of motion (EOMs) for the system must be determined. The EOMs for the momenta can be easily determined as for any system described using a Hamiltonian, and are determined by the forces on the beads and Fourier components.
\begin{equation}
  \frac{\partial \mathbf{p}_{i,j}}{\partial t} = - \frac{\partial H_{\mathrm{BF}}}{\partial\mathbf{q}_{i,j}}
   = \mathbf{F}_{i,j}
  ;\quad 
  \frac{\partial \mathbf{p}_{i,jk}}{\partial t} = - \frac{\partial H_{\mathrm{BF}}}{\partial\mathbf{a}_{i,jk}}
   = \mathbf{F}_{i,jk}.
  \label{eq:EOM-mom}
\end{equation}
The EOMs for the bead positions and Fourier amplitudes are related simply to their respective momenta,
\begin{equation}
  \frac{\partial \mathbf{q}_{i,j}}{\partial t} = \frac{\partial H_{\mathrm{BF}}}{\partial\mathbf{p}_{i,j}}
  = \frac{\mathbf{p}_{i,j}}{m_{i}}
  ;\quad 
  \frac{\partial \mathbf{a}_{i,jk}}{\partial t} = \frac{\partial H_{\mathrm{BF}}}{\partial\mathbf{p}_{i,jk}}
  = \frac{\mathbf{p}_{i,jk}}{m_{i,k}}
  \label{eq:EOM-pos}.
\end{equation}

To perform dynamics in the canonical ensemble for calculating thermal averages, the system must be coupled to a thermostat to maintain the desired temperature. For our work, we make use of a thermostat that has been designed for use in PI simulations, namely the PI Langevin equation (PILE) thermostat \cite{sepCeriotti2010}. This stochastic thermostat is very simple to implement, efficient, and only requires a single parameter. The thermostat updates the momenta as,
\begin{equation}
  p \leftarrow c_{1}p + \sqrt{ \frac{m}{\beta}} c_{2}\zeta,
  \label{eq:pile}
\end{equation}
where $\zeta$ is a Gaussian random number with zero mean and unit variance. 
The constants $c_{1}$ and $c_{2}$ are defined as,
\begin{equation}
  c_{1} = e^{-(\Delta t /2)\gamma};\quad c_{2} = \sqrt{1-c_{1}^2},
  \label{eq:pile-consts}
\end{equation}
where $\Delta t$ is the simulation time step. The friction coefficients, $\gamma$, for BF-PIMD are dependent on which degrees of freedom (DOFs) they are acting on. The bead momenta are thermostatted in their normal modes, with $\gamma$ proportional to the frequency of the normal mode, except for the centroid mode, where it is the inverse of the time constant $\tau_{0}$, the one input parameter. The friction coefficients for the momenta for Fourier component $k$ are equal to $2 \omega_{n}k\pi/\sqrt{2} $.

BF-RPMD is performed in a similar manner as BF-PIMD but without a thermostat being applied to the bead momenta. In Paper II\cite{BF2} we showed that including a thermostat of the Fourier components was required to create the correct dynamics of our model systems. For atomistic BF-RPMD, we critically damp the Fourier components; thus, we use the same friction coefficients as used in BF-PIMD.

\section{Computational Details}\label{sec:Details}

\subsection{Implementation into DL\_POLY Quantum for atomistic simulations}

The BF-PI framework poses some challenges when implemented in atomistic simulations. Mainly, this is due to the fact that the Hamiltonian in Eq.~\ref{eq:BF-Ham}, and therefore the forces in Eq.~\ref{eq:EOM-mom}, contain an integral over the path between beads. The accuracy of this integral directly determines the accuracy of the dynamics. While for a simple model system like a harmonic oscillator these integrals can be solved analytically,\cite{junIvanov2003} for a general system they must be evaluated numerically.

Previous BF-PIMD studies \cite{junIvanov2003} (and Papers I\cite{octLondon2025} and II\cite{BF2}) have made use of the trapezoid rule to perform these integrals. The trapezoid rule approximates a definite integral using the sum of the areas of trapezoids with equal width,
\begin{equation}
  \int_{a}^{b} \mathrm{d}x\ f(x) \approx \sum_{l=1}^{l_{\mathrm{max}}} \frac{1}{2}\left(f(x_{l})+f(x_{l+1})\right)\Delta
  x,
  \label{eq:trapezoid}
\end{equation}
where $\Delta x = x_{l+1} - x_{l}$. A trapezoid rule calculation with $l_{\mathrm{max}}$ trapezoids will require $l_{\mathrm{max}}+1$ function evaluations. For BF-PI simulations, those function evaluations would involve invoking the force engine to calculate the gradients for the EOMs. Previous model system studies have shown that it takes around 10 segments to converge these integrals \cite{junIvanov2003,octLondon2025}. Other BF-PIMD studies have used the Simpson rule for integration\cite{julIvanov2005}, which also required a large number of points to converge. This large number of force evaluations is not tractable for atomistic simulations and would make the method more computationally expensive than B-PIs.

To circumvent this, we explored various integration schemes. One highly efficient method for numerical integration is using Gaussian quadrature.\cite{Press2007} In general, the integral for some function, $f(x)$, multiplied by a weighting function, $ W(x) $, can be approximated as,
\begin{equation}
  \int_{a}^{b} \mathrm{d}x\ W(x)f(x) \approx \sum_{l=1}^{l_{\mathrm{max}}} w_{l}f(x_{l}).
  \label{eq:guass}
\end{equation}
The set of integration points, $x_{l}$, is no longer necessarily an evenly spaced grid like in the trapezoid rule, but instead is determined by a set of orthogonal polynomials \cite{Press2007}. The set of polynomials is determined by the form of the weighting function. The case where $W(x)=1$ results in Gauss-Legendre quadrature (GLQ), where the set of 
$\left\{x_{l}\right\}$ is the roots of the Legendre polynomial of order $l_{\mathrm{max}}$, $P_{l_{\mathrm{max}}}(x)$. The weights, $\{w_{l}\}$, are found based on the derivatives of the Legendre polynomial \cite{Press2007},
\begin{equation}
  w_{l} = \frac{2}{(1-x_{l}^2)\left(P^{'}_{l_{\mathrm{max}}}(x_{l})\right)^2}.
  \label{eq:weights}
\end{equation}
The standard GLQ is for an integral over the interval $ \left[-1,1\right] $, but can be easily
adjusted to the general interval $ \left[a,b\right] $. The number of integration points, $ l_{\mathrm{max}} $, is determined by the form of the integrand function, where the integral is exact for a polynomial of order $2l_{\mathrm{max}}-1$ \cite{Press2007}. If the function is well approximated by a polynomial, GLQ can be an efficient way to perform a numerical integral of that function.

We have implemented BF-PIMD and BF-RPMD into our in-house software DL\_POLY Quantum\cite{aprLondon2024,mayLondon2025} with a new module added to handle the BF-PI framework. In particular, we included subroutines to calculate and store the function points and weights for GLQ at the start of the simulation. For a given simulation step, the software will calculate the BF path values and pass them to its standard force engine. The integration of the EOMs then follows the same velocity-Verlet scheme as used for B-PIs. The Fourier amplitudes are written out to their own trajectory file during the simulation. If the user is looking to calculate bead-estimated energies during the simulation, an additional set of calls to the force engine is performed periodically using the bead positions.

\subsection{Simulation Details}

\begin{figure*}
  \centering
  \includegraphics[width=0.99\textwidth]{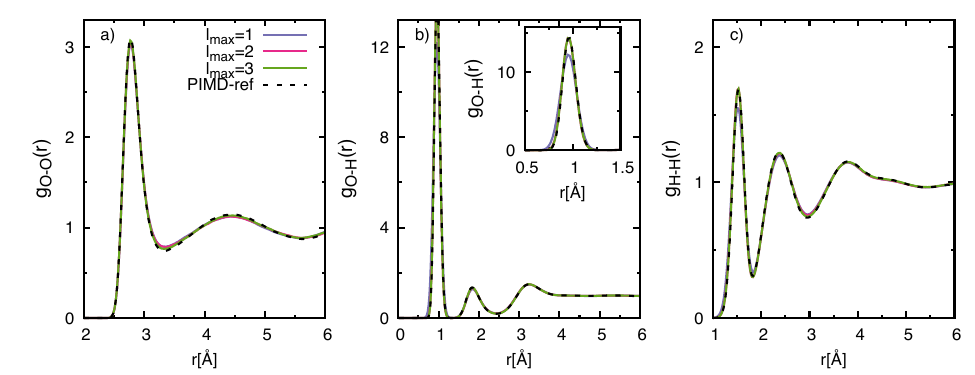}
  \caption{Calculated (a) oxygen-oxygen, (b) oxygen-hydrogen, and (c) hydrogen-hydrogen RDFs for liquid bulk water at 300 K with 216 q-TIP4P/F water molecules using BF-PIMD with $n=8$, $k_{\mathrm{max}}=1$, and $l_{\mathrm{max}}=1-3$. Reference B-PIMD with $n=32$ is shown in black dashed lines.}
  \label{fig:rdf-water}
\end{figure*}
  Simulations of liquid bulk water were performed using a box of 216 water molecules at 300 K described using the q-TIP4P/F potential.\cite{julHabershon2009} The water molecules were placed in a cubic box of length 18.64 \AA\ and periodic boundary conditions to achieve a density of 0.997 g/cm $ ^{3} $. Simulations of hexagonal ice were performed using 96 q-TIP4P/F water molecules at 150 K in an orthorhombic box with dimensions 15.65 \AA\ $\times$ 13.56 \AA\ $\times$ 14.76 \AA. 

Canonical simulations for both B-PIMD and BF-PIMD were performed using a time step of 0.2 fs. Simulations of 110 ps were performed, with the first 10 ps discarded as equilibration and the remaining 100 ps used to calculate statistical quantities and to sample initial conditions for microcanonical simulations.

To study the convergence of atomistic BF-PIMD for liquid bulk water, we ran simulations using $n=8$ beads with varying numbers of Fourier components ($k_{max}=1-3$) and integration points ($l_{max}=1-3$). B-PIMD reference simulations were performed using
  $n=32$. For hexagonal ice, we performed similar benchmarks, except adding $n=16$ for BF-PIMD and $n=64$ for the reference B-PIMD.

BF-RPMD simulations were performed for liquid bulk water using the converged $k_{max}$ and $l_{max}$ parameters. A total of 100 trajectories were performed to calculate the dynamical properties, with the initial configurations being sampled every 2 ps from the BF-PIMD simulations performed previously. BF-RPMD trajectories were run for 22 ps, with the last 10 ps being used for property calculations. The trajectories were split into groups of 20 to compute standard deviations of the calculated properties. Reference B-RPMD simulations were performed in a similar manner.

The diffusion coefficient was calculated using the center-of-mass velocity autocorrelation function,
\begin{equation}
  D = \frac{1}{3}\int_0^{\infty}\ \mathrm{d}t c_{\mathrm{vv}}(t),
\end{equation}
with the numerical integral being taken over 2 ps.

  Orientation relaxation times were calculated from orientation time correlation functions (TCFs). Here, we consider only the orientation relaxation along the OH bond unit vector, $\hat{\mathbf{u}}_{\mathrm{OH}}$. TCFs of various orders can be calculated as,
  \begin{equation}
    C_{l,\mathrm{OH}} = \left\langle P_{l}\left[ \hat{\mathbf{u}}_{\mathrm{OH}}(0)
      \hat{\mathbf{u}}_{\mathrm{OH}}(t)\right]\right\rangle
    \label{eq:TCF}
  \end{equation}

  The orientation relaxation times ($\tau_{l}$) are then calculated by fitting the TCFs to a biexponential function and then performing a weighted average of the two exponential constants.\cite{novShi2023}
  
All bead-Fourier observables reported in this work are calculated using the bead-estimator, where only the bead positions are used in the calculation instead of the full path. A more detailed study on calculating BF quantities has been explored in a previous work\cite{junIvanov2003} as well as in both Papers I\cite{octLondon2025} and II\cite{BF2}.

  \section{Results and Discussion}\label{sec:Results}

\subsection{Radial Distribution Functions}

  To study the performance of atomistic BF-PIMD simulations in calculating thermal equilibrium properties, we examine the radial distribution functions (RDFs) of liquid bulk water at 300 K and ice at 150 K as test cases. The full convergence plots for BF-PIMD with various numbers of Fourier components and integration points are given in the Supplementary Material. 

  Figure \ref{fig:rdf-water} shows the BF-PIMD calculated RDFs for liquid bulk water with $n=8$ and $ k_{\mathrm{max}}=1$ with varying numbers of integration points compared to the reference B-PIMD with $n=32$. For the oxygen-oxygen interactions shown in panel a, all values of $l_\mathrm{max}$ give excellent agreement with the reference B-PIMD simulations. 
  \begin{figure*}
  \centering
  \includegraphics[width=0.99\textwidth]{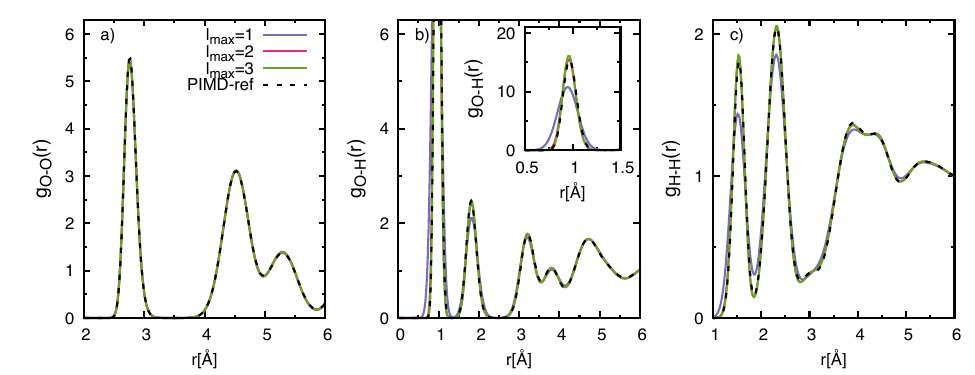}
  \caption{ Calculated (a) oxygen-oxygen, (b) oxygen-hydrogen, and (c) hydrogen-hydrogen RDFs for hexagonal ice at 150 K with 96 q-TIP4P/F water molecules using BF-PIMD with $n=8$, $k_{\mathrm{max}}=1$, and $l_{\mathrm{max}}=1-3$. Reference B-PIMD with $n=64$ is shown in black dashed lines.}
  \label{fig:rdf-ice}
\end{figure*}  
However, the interactions involving hydrogens are clearly not accurately captured by a single integration point. It is thus necessary to use two integration points for these simulations, which gives excellent agreement with the reference simulations for all three RDFs. While the second integration point requires an additional force call per bead, adding one Fourier component has resulted in a 4-fold reduction in the number of beads, so there are still fewer force calls than in the B-PIMD simulations. More discussion of the computational performance of BF-PIMD compared to B-PIMD is given in Sec. \ref{sec:performance}

We also explore the performance of BF-PIMD for the hexagonal ice system at 150 K, where the lower temperature and the increased role of NQEs result in the reference B-PIMD simulations requiring 64 beads, twice that of the liquid bulk water system. Fig.~\ref{fig:rdf-ice} showcases the BF-PIMD calculated RDFs for $n=8$ and $k_\mathrm{max}=1$, the converged parameters for this system. The full convergence plots, including BF-PIMD with 16 beads, are given in the Supplementary Material. As with the liquid bulk water system, it is necessary to include two integration points to converge the BF-PIMD integrals for one Fourier component. We do note that for larger numbers of Fourier components when using 8 beads, it is necessary to go beyond two integration points for convergence. This is not the case for $k_\mathrm{max}=1$ or for any case when using 16 beads. The increased number of integration points is likely due to the fact that, for a fixed number of beads, the length of the imaginary time path between them increases in length. That, coupled with the increased complexity of the path when introducing more terms to the Fourier sine series, can necessitate more terms for the approximation of the Gaussian quadrature to be accurate.

The performance of BF-PIMD for the hexagonal ice system is even more impressive than for liquid bulk water, with an 8-fold reduction in number of beads with the addition of a single Fourier component. Again, a more complete discussion of the computational performance will be done in Sec.~\ref{sec:performance}.

\subsection{Dynamical Properties}
 
  \begin{figure}
    \centering
    \includegraphics[width=0.99\linewidth]{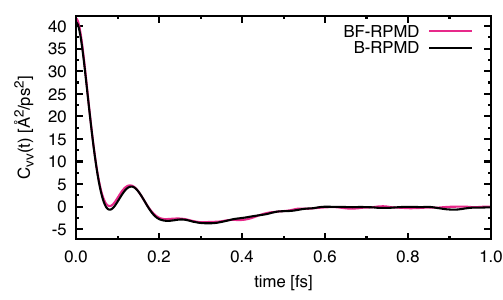}
    \caption{Calculated velocity autocorrelation function of 216 q-TIP4P/F liquid bulk water at 300 K using (pink) BF-RPMD with $n=8$, $k_\mathrm{max}=1$, and $l_\mathrm{max}=2$ and (black) B-RPMD with $n=32$.}
    \label{fig:corr}
  \end{figure}

  \begin{figure}
    \centering
    \includegraphics[width=0.99\linewidth]{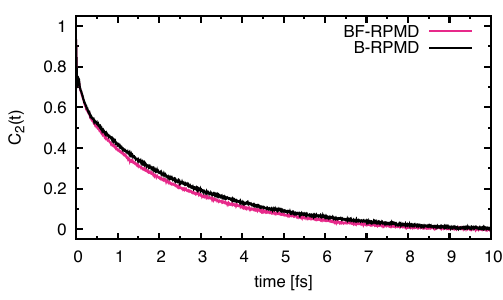}
    \caption{Calculated second-order orientation autocorrelation function of 216 q-TIP4P/F liquid bulk water at 300 K using (pink) BF-RPMD with $n=8$, $k_\mathrm{max}=1$, and $l_\mathrm{max}=2$ and (black) B-RPMD reference with $n=32$}
    \label{fig:orientation}
  \end{figure}

  \begin{table}
    \caption{Calculated dynamical properties of 216 q-TIP4P/F liquid bulk water at 300 K, comparing BF-RPMD with $n=8$, $k_{\max}=1$, $l_{\max}=2$, B-RPMD with $n=32$, classical MD, and experiment.}
    \begin{tabular}{c c c c c }
      & BF-RPMD & B-RPMD & Classical & Exp.\\
      \hline
      D [\AA$^2$/ps] & 0.25 $ \pm $ 0.02 & 0.21 $\pm$ 0.04 & 0.17 $\pm$ 0.03 & 0.229\cite{janKrynicki1978} \\
      $ \tau_{1} $ [ps]  & 4.4 $\pm$ 0.2 &  4.9 $\pm$ 0.3 & 5.2 $\pm$ 0.2 & - \\
      $ \tau_{2} $ [ps]  & 1.9 $\pm$ 0.2 & 2.1 $\pm$ 0.2 & 2.55 $\pm$ 0.09  & 1.7-2.6\cite{sepWinkler2000,janLawrence2003,sepRezus2005} \\
       \hline
    \end{tabular}
    \label{tab:dyn}
  \end{table}
  
One of the key strengths of the BF-PI framework is that it can be directly applied to real-time dynamical simulations without any ad hoc approximations. Here, we present BF-RPMD calculated dynamical properties of liquid bulk water at 300 K as a test case. These simulations are performed with the converged parameters obtained from BF-PIMD, namely $n=8$, $k_{\mathrm{max}}=1$, and $l_{\mathrm{max}}=2$.

The calculated velocity ACFs are presented in Fig.~\ref{fig:corr}, for which we see good agreement between BF-RPMD and B-RPMD as expected. We also observe a good agreement in the calculated second-order orientation ACFs as shown in Fig.~\ref{fig:orientation}. All calculated dynamical properties are tabulated in Table~\ref{tab:dyn}. These results are consistent with the results of Paper II\cite{BF2}, which showed that BF-RPMD is able to capture the same approximate real-time quantum dynamics as B-RPMD while being able to reduce the number of beads needed to converge. In this work, we demonstrate that the method can also be used for condensed-phase atomistic simulations, the first time BF-PIs have been used in this context.

We note that, for these BF-RPMD results, no ad hoc corrections are applied or needed. The only alterations from a pure BF-RPMD implementation are that we have included a thermostat on the Fourier amplitudes, as was done previously for the model systems in Paper II\cite{BF2}. This gives us confidence that BF-RPMD is well-suited to any atomistic simulations for which B-RPMD works.

  \subsection{Computational Performance}\label{sec:performance}

\begin{figure}
  \centering
  \includegraphics[width=0.99\linewidth]{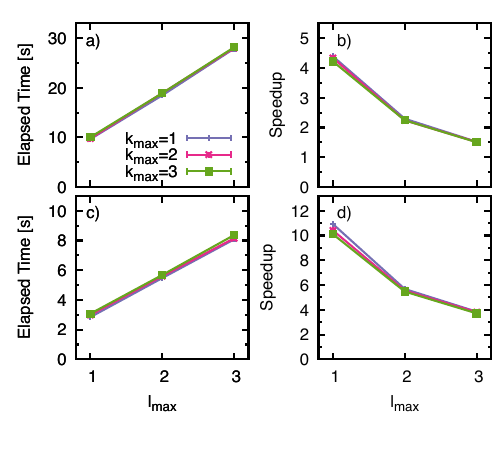}
  \caption{a) Time to complete 100 time steps for BF-PIMD simulations of 216 q-TIP4P/F water molecules with $n=8$ for varying numbers of Fourier components ($k_{\mathrm{max}}$) and integration points ($l_{\mathrm{max}}$) averaged over 3 separate simulations. Simulations were performed in serial on the RM nodes of the Bridges2 computing cluster\cite{access}. b) Speedup of BF-PIMD simulations compared to timings of 32-bead B-PIMD simulations. c-d) Similar to a) and b) but for hexagonal ice using 96 q-TIP4P/F water molecules and B-PIMD reference with 64 beads. Error bars are within the size of the points.}
  \label{fig:8bd-timing}
\end{figure}

While it is useful to know that BF-PIMD and BF-RPMD yield results consistent with their standard bead counterparts, one of the primary goals of the method is to reduce the computational cost of the simulations. In this section, we analyze the performance of BF-PI methods in terms of the simulation runtime. To evaluate the performance of the methods on equal footing, we performed a series of short simulations for both the liquid bulk water and hexagonal ice systems using BF-PIMD with 8 beads and B-PIMD with 32 beads and 64 beads for the respective systems. All simulations were performed in serial using DL\_POLY Quantum on a full RM node of the Bridges 2 cluster\cite{julBrown2021} to avoid conflicts with other jobs. For each set of $k_\mathrm{max}$ and $l_\mathrm{max}$, three independent simulations were run and averaged over. Simulations were run for 1100 time steps, with the time to complete 100 time steps averaged over the last 1000 time steps.

Panels a) and c) of Fig.~\ref{fig:8bd-timing} show the time to complete 100 time steps for both liquid bulk water and hexagonal ice. As expected, we observe a linear increase in the simulation runtime with increasing the number of integration points, since we are increasing the number of calls to the force engine. The more useful comparison is the speedup relative to B-PIMD, as shown in panels b) and d). While the largest speedup is, of course, seen for simulations using only one integration point, they do not yield converged results, as discussed previously. The speedup observed for two integration points remains pronounced and is even larger than expected when focusing only on the force evaluations. 

To understand the observed speedups, it is important to analyze the number of calls to the force engine. For B-PIMD, every time step requires $n$ calls to the force engine, while for BF-PIMD it is $n l_\mathrm{max}$. The number of Fourier components does not directly affect this, as the forces are evaluated along the path, which has already performed the sum over Fourier components. However, the number of Fourier components can affect the needed $l_\mathrm{max}$ to converge the simulations depending on the system, as discussed previously.

For the case of our converged data for the liquid bulk water (i.e., $n=8, k_\mathrm{max}=1, l_\mathrm{max}=2$), we obtain a speedup of $\sim$2.25x compared to B-PIMD, which is larger than the expected 2x. The difference is even larger for hexagonal ice, where our converged data using the same parameters should yield a 4-fold speedup, while we see a $\sim$5.5x speedup.

These results indicate that the improved performance of BF-PIMD over B-PIMD is due to more than just the reduction in the number of force evaluations. While the force evaluation is typically the most computationally demanding part of an MD simulation, BF-PIMD has also reduced the cost of other parts of the simulation as well. This includes reducing the number of neighbor list calculations by reducing the number of beads for non-bonded interactions. For our implementation of BF-PIMD in DL\_POLY Quantum, we calculate the neighbor list for the same points along the path on which we perform the force evaluations, resulting in the same reduction in neighbor list calculations as in force evaluations. Additionally, both BF-PIMD and B-PIMD make use of normal modes to transform the free ring polymer into $n$ independent harmonic oscillators. However, there is a need to constantly transform the normal mode coordinates into Cartesian ones to be able to calculate the forces from the external potential. While this is a simple process, reducing the number of beads can greatly improve simulation performance. This is likely why we obtain a more dramatic speedup for the hexagonal ice system than for the liquid water system, as our bead-reduction factor is even greater. As such, BF-PIMD is at its strongest for systems with more significant NQEs, where it can potentially achieve larger reductions in the number of beads than B-PIMD.

While these performance metrics were calculated for BF-PIMD, the reductions in time to perform a time step will, of course, carry over to the real-time BF-RPMD simulations. This will also make calculations of dynamical properties more efficient, increasing and extending the impact of the computational savings from using BF-PIMD to BF-RPMD.

  \section{Conclusion}\label{sec:Conclusion}

  In this work, we have demonstrated the first implementation and application of BF-PIs in atomistic condensed-phase simulations, utilizing BF-PIMD for both liquid water and hexagonal ice and BF-RPMD for liquid water. In doing so, we have obtained up to an 8-fold reduction in the number of beads needed to calculate thermal equilibrium properties of hexagonal ice and a 4-fold reduction in the number of beads needed to calculate dynamical properties of liquid bulk water by adding a single Fourier component. This is consistent with our previous results for model systems. We also estimate that BF-RPMD could offer speedups of up to 5.5x over traditional B-RPMD methods for systems with prominent NQEs.

  Future work is to continue to explore how BF-PIs can be used in atomistic simulations. A natural extension is to implement a BF-TRPMD to calculate vibrational spectra. Additionally, we will explore using BF-PIs to incorporate NQEs into simulations with higher-accuracy forces, whether in AIMD or in dynamics with HDNNPs\cite{behler2007generalized, faruque2026predictive}, where the large number of beads used in traditional B-PIs may become computationally demanding.

  \section*{Supplementary Material}
The supplementary material contains convergence details for the Fourier components and the number of beads for the bulk water and hexagonal ice systems.

\begin{acknowledgments}
Simulations presented in this work used resources from Bridges-2 at Pittsburgh Supercomputing Center through allocations PHY230030P and CHE240103 from the Advanced Cyberinfrastructure Coordination Ecosystem: Services \& Support (ACCESS) program,\cite{access} which is supported by National Science Foundation grants \#2138259, \#2138286, \#2138307, \#2137603, and \#2138296. The authors also acknowledge the HPC center at UMKC for providing computing resources and support.
\end{acknowledgments}

\subsection*{Conflict of Interest}

\noindent The authors have no conflicts to disclose.

\subsection*{Author Contributions}
\textbf{Nathan London}: Method (equal); Writing - original draft (lead); Writing - review \& editing (equal). 
\textbf{Mohammad R. Momeni}: Conceptualization (lead); Funding acquisition (lead); Supervision (lead); Method (equal); Writing - original draft (supporting); Writing - review \& editing (equal). 

\section*{\label{sec_contrib} DATA AVAILABILITY}
\noindent  The data that support the findings of this study are provided in the main text and the Supplementary Material. Additional data are available from the corresponding author upon request.

\section*{REFERENCES}

\bibliography{bib}

\end{document}